\documentclass[aps,prb,floatfix,twocolumn]{revtex4}
\usepackage{graphicx}
\usepackage{amsmath}
\usepackage{amssymb}

\newcommand{\BEQ}{\begin{eqnarray}}
\newcommand{\EEQ}{\end{eqnarray}}
\newcommand{\BEA}{\begin{eqnarray}}
\newcommand{\EEA}{\end{eqnarray}}
\newcommand{\nn}{\nonumber}
\renewcommand{\d}{{\rm d}}

\newcommand{\tauc}{{\tau_{\rm collapse}}}
\newcommand{\taur}{{\tau_{\rm reg}}}

\newcommand{\tr}{{\rm tr}}

\newcommand{\HH}{\hat{H}}

\newcommand{\RM}{{\rm M}}
\newcommand{\RB}{{\rm B}} 
\newcommand{\RSA}{{\rm SA}}
\newcommand{\down}{{\downarrow}}
\newcommand{\up}{{\uparrow}}
\newcommand{\uu}{{\uparrow\uparrow}}
\newcommand{\dd}{{\downarrow\downarrow}}
\newcommand{\ud}{{\uparrow\downarrow}}
\newcommand{\du}{{\downarrow\uparrow}}

\newcommand{\ri}{{\rm i}}
\newcommand{\tf}{t_{\rm f}}

\newcommand{\um}{{\underline m}}
\newcommand{\mc}{{m_{\rm c}}}
\newcommand{\half}{\frac{1}{2}}

\newcommand{\CD}{{\cal D}}
\begin{document} 
\title
{The quantum measurement process: an exactly solvable model}
\author{A.E. Allahverdyan$^{1,2)}$, R. Balian$^{3)}$
and \underline{Th.M. Nieuwenhuizen}$^{1)}$\footnote{The author who 
presented this contribution.}}
\affiliation{$^{1)}$ Institute for Theoretical Physics,
Valckenierstraat 65, 1018 XE Amsterdam, The Netherlands}
\affiliation{$^{2)}$Yerevan Physics Institute,
Alikhanian Brothers St. 2, Yerevan 375036, Armenia}
\affiliation{$^{3)}$ SPhT, CEA-Saclay, 91191 Gif-sur-Yvette cedex, France}

\begin{abstract}
An exactly solvable model for a quantum measurement is discussed, that
integrates quantum measurements with classical measurements.

The $z$-component of a spin-$\half$ test spin is measured with an apparatus, 
that itself consists of magnet of $N$ spin-$\half$ particles, coupled to a bath.
The initial state of the magnet is a metastable paramagnet, while the bath
starts in a thermal, gibbsian state. 
Conditions are such that the act of measurement drives the magnet in the 
up or down ferromagnetic state, according to the sign of  $s_z$ of 
the test spin. 

The quantum measurement goes in two steps.
On a timescale $1/\sqrt{N}$ the collapse takes place due to a unitary
evolution of test spin and apparatus spins; on a larger but
still short timescale this collapse is made definite by the bath.
Then the system is in a `classical' state,
having a diagonal density matrix. The registration of that state is
basically a classical process, that can already be understood 
from classical statistical mechanics.
\end{abstract}
\maketitle

Quantum mechanics has filled last century with respect and debate. 
Respect for its accurate description of e.g. the solid state, 
quantum chemistry, high energy physics and the early universe.
Debate has remained about its foundations: 
what exactly is this theory standing for? 

One of the most fundamental questions~\cite{wh}  is still: 
how should we describe and understand a quantum measurement?
The first answer this  question came from Bohr, who stated that
the apparatus should be classical. This separation between
classical world versus quantum world
(apparatus versus testsystem) was felt to be unnatural. 
von Neumann considered an apparatus as
macroscopic and postulated that it induces a
collapse of the wavefunction (reduction of the wavepacket),
so his complete quantum mechanics consists of
Schr\"odinger dynamics plus the collapse postulate.
Here a similar separation has been introduced, namely macroscopic
versus microscopic.

To investigate the matter, several models have been 
proposed~\cite{models}, which did not converge to a unique picture.
Here we discuss here an exactly solvable model~\cite{ABNEPL}, 
from which the general structure, found before in a more
complicated bosonic model~\cite{ABNBose}, can be read off. 
As foreseen on general grounds~\cite{vKampen},~\cite{balian},
the measurement takes place in two steps: 
on timescale $\tauc\ll \hbar/T $, the quantum timescale,
a collapse of the wavefunction (reduction of the wavepacket) 
takes place,  while on a timescale $\taur\gg \hbar/T$ the registration 
of the measurement occurs. 
We shall discuss that the registration just coincides with a 
measurement of a `classical' Ising spin $s_z=\pm 1$ with the same apparatus.

Our solution for the measurement problem is compatible with the 
statistical interpretation of quantum mechanics and rules 
out several competing interpretations.

\subsection*{\it Classical measurement} 

Since our approach aims to give a unified view of quantum and classical 
measurements,  let us first recall how to measure a classical 
Ising spin (classical two-state system), which is in a definite state 
$s_z=\pm 1$. It is known that some classical systems may indeed
be approximately described as a two-state object, e.g. a classical 
brownian particle in a double-well potential with well-separated minima 
and a steep potential barrier in between. 

Our apparatus (A) consists of a magnet (M) coupled to a bath (B). 
The magnet contains N Ising spins $\sigma_z^{(n)}=\pm 1$ having a 
mean-field interaction between all quartets 

$$H_\RM = -\frac{J}{4 N^3} \sum_{ijkl=1}^N 
\sigma_z^{(i)}\sigma_z^{(j)}\sigma_z^{(k)}\sigma_z^{(l)}
= -\frac{1}{4}NJ{\underline m}^4,\eqno(1)$$         
with ${\underline m}=(1/N)\sum_n\sigma_z^{(n)}$ the fluctuating
magnetization. In the standard Curie-Weiss model all pairs would be
coupled, and a second order phase transition occurs.
The quartic interaction has been chosen in order to have a first order
transition, as it happens in a bubble chamber, where an oversaturated
gas goes into droplets of its stable phase, the liquid,
when triggered by a particle.

The interaction between the test system S and the apparatus
$$ H_\RSA=-gs_z\sum_n\sigma_z^{(n)}=-gs_zN{\underline m},\eqno (2)$$
is turned on at $t=0$, the beginning of the measurement, and turned off
at the final time $t_{\rm f}$.

Initially the magnet starts in the paramagnetic state: each spin has 
chance $\half$ to be up or down, implying a zero average magnetization,
$m\equiv\langle\um\rangle=0$.

At a critical temperature $T_c$ the magnet undergoes  a phase transition
to a state with magnetization $\pm \mc$.
Due to the quartic interactions (2), it is a first order phase transition.
The free energy per spin, $F=U-TS$, reads

$$ F=- \frac{J m^4}{4}-g s_zm-
T(\frac{1+m}{2}\ln\frac{2}{1+m}+\frac{1-m}{2}\ln\frac{2}{1-m}),
\eqno (3) $$ 
At $g=0$ and for $T$ below $T_c=0.362949\,J$, the paramagnet $m=0$ is 
still metastable, see  Fig. 1. 
It is here that the setup lends itself as an apparatus for a measurement:
by starting in the metastable paramagnet one has the magnetic
analog of the metastable oversaturated gas of a bubble chamber.

\begin{figure}
\includegraphics[width=7cm,height=5cm]{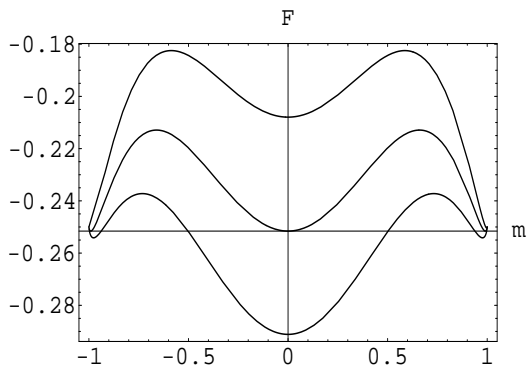}
\caption{Free energy of the magnet as function of $m$.
Lower curve: at large $T=0.42\,J$ the paramagnet $m=0$ has lowest free energy.
Middle curve: at $T_c=0.362949\,J$ the local minima 
$m=0$, $m_\up=0.990611$ and $m_\down=-m_\up$ become degenerate.
Upper curve: Below $T_c$, here  $T=0.3\,J$, the paramagnet is metastable, 
while the minima $m_\up$, $m_\down$ are stable. 
In the measurement the magnet start in the metastable state 
and ends up in one of the stable states.}
\end{figure}

\begin{figure}
\includegraphics[width=7cm,height=5cm]{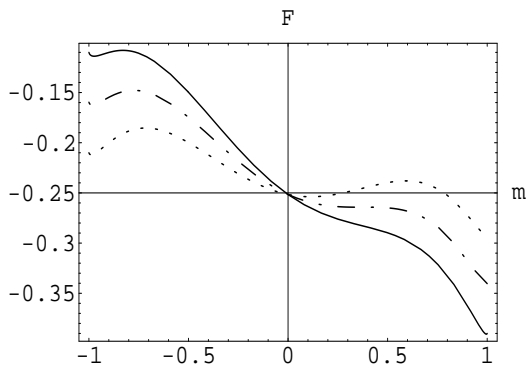}
\caption{A free energy barrier can be overcome 
by the coupling. Here $T=T_c$ and $s_z=+1$. 
Dotted curve: the small coupling $g=0.04\,J$ does not suppress the barriers.
The setup cannot bring the magnetization from $m=0$ to the minimum near $m=1$. 
Dash-dotted curve: at the critical value $g_c= 0.09035\,J$  the barrier 
near $m=0.5$ is just suppressed. 
Full curve: at large coupling, $g=0.12\,J$, there is no barrier
and $m$ will end up in the minimum to register the measurement.
For $s_z=-1$ the left barrier would be suppressed.}
\end{figure}

At time $t=0^+$ the coupling $g$ between the test-spin and the apparatus
 is turned on, which puts the magnet in an external field $gs_z=\pm g$, see Eq. (2). 
If $g$ is large enough and $s_z=+1$, the interaction suppresses the barrier 
near $m=0.7$, see Figure 2, and for $s_z=-1$ it will suppress the one near $m=-0.7$.
This is the magnetic analog of a bubble in a bubble chamber, where a supercritical 
gas is triggered to bubbles of its liquid state by a test particle.

Let us denote by $\up$ and $\down$ the $s_z=\pm1$ cases. With the field 
turned on, the magnetization will move from $m=0$ to the
minimum of $F$. This is possible due to a weak coupling to the bath,
which allows to dump the excess energy in the bath. 
The dynamics of $m$ was coined on the basis of detailed
 balance alone~\cite{SuzKubo}, but that is not enough to fix it. 
We consider the model where all three spin components of all $N$ apparatus spins
are weakly coupled to independent Ohmic bosonic baths 
(sets of harmonic oscillators). The proper dynamics appears to be: 
$$\dot{m}=
\gamma h (1-\frac{m}{\tanh\beta h}),\quad h=gms_z+Jm^3, \eqno (4)$$

where $\gamma\ll 1/\hbar$ is a small parameter charactering the 
weak coupling to the bath.  For $s_z=+1$,  $m$ will go to the right in Fig. (2).
After the $m$ has approached the minimum $m_\ast$, 
the apparatus is decoupled, $(g\to 0)$.
The magnetization will then move from $m_\ast$ to the $g=0$ - minimum 
$m_\up\approx m_\ast-0.004$. 
It will stay there up to a hopping time $\exp(N)$; 
for large $N$ this means ``for ever''. Whether or when the apparatus is 
read off (``observation'') is immaterial.  

The measurement has been performed: if $s_z$ was $+1$, the apparatus has 
ended up with magnetization $+m_\up$, and for $s_z=-1$ the magnetization 
went to $m_\down=-m_\up$.

Repeating the measurement will then allow to determine an ensemble with spins
having $s_z=1$ with probability $p_\up$ and $s_z=-1$ with probability 
$p_\down=1-p_\up$.

\subsubsection*{ The statistical interpretation of quantum mechanics.}

We can describe the quantum measurement by adopting the statistical 
interpretation put forward by Einstein, see e.g. ~\cite{ballentine},
~\cite{balian}. The most important aspects are:
1) A quantum state is described by a density matrix.
2) A single system does not have "its own" density matrix or wavefunction;
3) Each quantum state describes an ensemble of identically prepared systems;
this also  holds for a pure state $|\psi\rangle\langle\psi|$.

In this approach a quantum measurement must describe  an ensemble of 
measurements on an ensemble of systems. The task is to show that all 
possible outcomes occur with Born probabilities and proper correlations
between test system and apparatus.

\subsubsection*{\it Quantum measurement.} 

In our model the above setup carries over immediately to the quantum 
situation. 
First, the Ising test spin $s_z$ should be replaced by the 2x2 Pauli matrix 
$\hat s_z$, and the apparatus spins $\sigma_z^{(n)}$ by  $\hat \sigma_z^{(n)}$.
The magnetization operator $\hat m=(1/N)\Sigma_n \hat \sigma_z^{(n)}$ will
enter the Hamiltonians (1) and (3). In the Hamiltonian of the bath
and the interaction Hamiltonian between the magnet spins and the bath,
there will occur creation and annihilation operators for the bosons,
while the interaction term has the Pauli operators for the spins.

The test spin may start in an unknown quantum state, that is to say, its
 spin-averages $\langle\hat s_x\rangle$, $\langle\hat s_y\rangle$ and 
$\langle\hat s_z\rangle$  are unknown and arbitrary; 
our measurement will determine the latter. 
On the basis where $\hat s_z$ is diagonal 
the initial density matrix $r(0)$ has the elements 
$r_{\uu}(0)=1-r\dd(0)=\half(1+\langle\hat s_z\rangle)$,   
$r_\ud(0)=r_\du^\ast(0)=\half(\langle\hat s_x\rangle-\ri\langle\hat 
s_y\rangle)$.

The full density matrix of the system $\CD$ reads initially:
$\CD(0)=r(0)\otimes R_\RM(0)\otimes R_\RB(0)$, where 
$R_\RM(0)=2^{-N}\Pi_n \hat\sigma_0^{(n)}$ describes
the paramagnet, so each spin is described by 
the identity matrix $(\hat\sigma_0^{(n)})_{ij}=\delta_{ij}$,
while $R_\RB(0)$ the equilibrium (Gibbs) state of the bath.
In particular, there is no initial correlation between test system and 
apparatus, in order to avoid any bias in the measurement. 

\subsubsection*{\it Selection of the collapse basis}

The dynamics is set by the von Neumann equation 
$\ri \hbar\frac{\d}{\d t} \CD=[\HH,\CD]$,
where $\HH$ is the full Hamiltonian operator, including also the bath and 
the coupling between magnet and bath.
The state of the test spin is $r(t)=\tr _{\RM,\RB} \CD(t)$.
For its evolution only the quantum version of the 
interaction Hamiltonian (3) remains, 

$$ \frac{\d}{\d t}\, r_{ij}  
= -gN(s_i-s_j)\,\tr_{\RM,\RB}[\,\hat{m},\CD_{ij} ], \eqno (5)$$
where $i,j=\up,\down$ and $s_\up=+1$,  $s_\down=-1$ are the eigenvalues 
of $\hat s_z$.
It follows that the diagonal elements are conserved in time: $r_\uu(t)=p_\up$,
$r_\dd(t)=p_\down$. This happens because the spin has no dynamics of its own.
This conservation is a sine-qua-non for a reliable measurement.
The off-diagonal elements are endangered, 
and actually will collapse.

We learn form this that the selection of the collapse basis is a direct 
consequence of forces exerted by the apparatus on the test system:
The choice of the interaction Hamiltonian sets the basis on which its 
system-sector  diagonalizes.  Zurek has claimed that the 
selection would be imposed by the coupling to the environment~\cite{Zurek},
even though the difficulties to control these couplings make it an 
undesired candidate for such an important issue. The above argument
does in no way invoke the environment and thus rules out Zurek's picture.

\subsubsection*{\it The collapse.} 
 
It will take place on a short timescale, where both the spin-spin 
interactions and the spin-bath interactions are still inactive. 
The problem is then simple: the evolution of $N$ independent 
apparatus spins, coupled to the test spin. 
For $\CD_\ud$ this means that each $ \hat\sigma_0^{(n)}$ = diag ($1,1$) 
evolves as diag ($e^{2\ri gt/\hbar}$, $e^{-2\ri gt/\hbar}$), implying

$$r_\ud(t)\equiv \tr_{\RM,\RB}\CD_\ud(t)
=r_\ud(0)[\cos\frac{2gt}{\hbar}]^N. \eqno(6)$$

For short times this exposes a Gaussian decay,
$r_\ud(0)\,\exp(-t^2/\tau^2_{\rm collapse})$, with collapse time

$$\tauc=\frac{\hbar}{g\sqrt{2N}}\ll \frac{\hbar}{T}. \eqno (7)$$

In the estimate we used that $g\sim J\sim T$ and $N\gg 1$.

The recurrent peaks of the cosines, at $t_k=k\pi\hbar/2g$,
are suppressed by the bath, which brings a factor 
$\sim \exp(-\gamma\hbar N)$, that vanishes when $N$ is large enough.
A small dispersion in the $g$'s is quite realistic,
and would bring an additional reduction by
$\exp(-k^2\pi^2\frac{\langle g^2\rangle-\langle g\rangle^2}
{2\langle g\rangle^2}\,N)$.

In conclusion, the collapse is a quantum coherent process on a short
timescale $1/\sqrt{N}$, agreeing with von Neumann's postulate if $N$ is
macroscopic. The recurrence of the peaks is suppressed on a later timescale, 
due to a small coupling of the macroscopic magnet to the bath (decoherence).
This is an effect of the `environment', but it is not the main cause
of the collapse.

\subsubsection*{\it Registration of the quantum measurement.} 

The off-diagonal sectors of the density matrix having decayed, it is time to 
study the diagonal ones. Now the bath and the coupling to its have to be 
specified in detail. 
For simplicity each spin of the magnet is assumed to have its own subbath. 
These subbaths are all identical but independent, consisting of 
harmonic oscillators 
in x,y and z-direction, which are coupled bilinearly to the components
of the spins, and start out in their Gibbs state. 
Working out this quantum problem we observe complete analogy to the above
 description termed ``classical measurement". In particular, the evolution 
of $m(t)$, Eq. (4) announced above, is derived form first principles. 
The characteristic timescale is much larger than the quantum time,

$$\taur=\frac{1}{\gamma g}\sim\frac{1}{\gamma J}\gg \frac{\hbar}{T}.
\eqno (8)$$

\subsubsection*{\it Result of the measurement.}
 
After the measurement, at $\tf\gg \taur$, the common state of test spin 
and apparatus is

\BEA \label{Dfin3}
D(\tf)&=&p_{\up}\times|\up\rangle\langle \up| \otimes 
\rho_{\up\up}^{(1)}(\tf)\otimes \cdots \otimes 
\rho_{\up\up}^{(N)}(\tf)\qquad\quad(9)  \nn\\ 
&+&  p_{\down}\times|\down\rangle\langle\down|\otimes
\rho_{\down\down}^{(1)}(\tf)\otimes\cdots\otimes 
\rho_{\down\down}^{(N)}(\tf)\nn. \EEA 
where $\rho_\uu^{(n)}(\tf)=\half\, {\rm diag}(1+m_\up,1-m_\up)$ 
is the Gibbs density matrix of spin $n$ of the magnet. 

In words: with probability $p_\up$ one 
finds the spin in the up-state, and the magnet with magnetization up, and
likewise in the down-sector. The off-diagonal sectors, called 
``Schr\"odinger cats", are eliminated by the collapse.

We can now turn off the apparatus (by setting $g=0$). The above
state will change slightly, because $m$ now goes to the $g=0$
minimum of the free energy, and the system will remain there forever.
Eq. (9) will hold in this modified form.

Let us stress the meaning of Eq. (9) according to the statistical
interpetation. In doing a series of experiments, there are two possible
outcomes, connected with the magentization of the apparatus being up or down,
which occur with probabilities $p_\up$ and $p_\down$, respectively.
In each such event, the $z$-component of the test spin is 
equal to $+1$ or $-1$, correspondingly. 
The ensemble of spins having $+1$ is described by
the pure state density matrix $|\up\rangle\langle \up|$, or simply
by the wavefunction $|\up\rangle$. A similar statement holds for the
down spins. 

Notice that within the Copenhagen interpretation one assumes that once a 
single event has happened, a single spin does have its own collapsed wavefunction. 

Another issue, namely whether single events can be accounted for by 
quantum mechanics, has to be answered negatively. Within the statistical
interpretation, quantum mechanics is a theory about the statistics of 
outcomes of many events.

\subsubsection*{\it Conclusion}

The initial paramagnetic state of the apparatus consists of many microstates, 
so a statistical description is called for, and we have 
retained that for the quantum measurement. This is 
possible within the statistical interpretation of quantum mechanics, which 
states that any quantum state describes an ensemble of systems.
A theory of quantum measurements must therefore
describe an ensemble of measurements on an ensemble of identically 
prepared systems.

It was found that the collapse occurs quite fast after the start of the 
measurement. It goes in two steps: the collapse occurs due to interaction 
of the test system with the macroscopic apparatus, 
and later is made definite by bath induced decoherence.

The registration of the measurement occurs in a ``classical'' state,
a state that has 
collapsed already. Here a naive classical approach and a detailed quantum 
approach yield exactly the same outcome. The pointer variable ends up in a 
stable thermodynamic state. Whether the outcome is observed or not is
 immaterial.

For a macroscopic apparatus the collapse will almost be instantaneous,
yielding the basis for the postulate of von Neumann.
Our theory can be tested by mapping out the $N$-dependence.

\subsubsection*{\it Other interpretations of quantum mechanics.}

Our approach makes some of the interpretations, that caused much dispute in
the past~\cite{wh}, obsolete. A multi-universe picture is incompatible, 
since a collapse does occur. 
Mind-body problems do not show up, because the act of observation is no more
than gathering information about the classical final state of the apparatus.
Not environment induced selection~\cite{Zurek} 
but the coupling to the apparatus is 
causing the collapse. Gravitation plays no role.
Extensions of quantum mechanics, like spontaneous or stochastic 
localization and spontaneous collapse models,  are not needed.

We find no support for interpretations that attribute a special role 
to pure states of test system, e.g. the modal interpretation
and Bohmian mechanics.
The assumption of an underlying pure state for the whole system is 
unnecessary and would anyhow be problematic for describing 
the statistical nature of the apparatus in realistic setups.
 
The work of A.E. A. is part of the research programme of the Stichting voor 
Fundamenteel Onderzoek der Materie (FOM), financially supported by 
the Nederlandse Organisatie voor Wetenschappelijk Onderzoek (NWO). 
R. B. is grateful for hospitality at the University of Amsterdam,
and A.E. A. and Th.M. N. for hospitality at the CEA Saclay.

\end{document}